\documentclass[twoside,fleqn]{article}
\usepackage{espcrc2,mathbbol}
\newcommand{\half}{{\scriptstyle{{1\over 2}}}}
\def\beq{\begin{equation}}
\def\eeq{\end{equation}}
\def\bea{\begin{array}}
\def\eea{\end{array}}
\def\beqa{\begin{eqnarray}}
\def\eeqa{\end{eqnarray}}
\def\pl{{{\cal P}_\infty}}

\def\cA{{\cal{A}}}
\def\tr{{\rm tr}} 
\def\Tr{{\rm Tr}} 
\def\cV{{\cal V}}
\def\cD{{\cal D}}
\def\k{{\bf k}}

\title{Calorons and fermion zero-modes
\vskip-3cm\hfill\small INLO-PUB-10/03; IFT-UAM/CSIC-03-29\vskip2.6cm
}
\author{Falk Bruckmann\address{Instituut-Lorentz for Theoretical Physics, 
University of Leiden,P.O. Box 9506, NL-2300 RA Leiden, The Netherlands.},
Margarita Garc\'{\i}a P\'erez\address{Instituto de F\'{\i}sica Te\'orica, 
C-XVI, Universidad Aut\'onoma de Madrid, Madrid 28049, Spain.},
D\'aniel N\'ogr\'adi${}^{\rm a}{}$ and Pierre van Baal${}^{\rm a}{}$
\thanks{Presented by the last author at Lat.'03, Tsukuba, Japan.}
}
\begin{document}
\begin{abstract}
Calorons in the confined phase for SU($n$) gauge theory, having a 
non-trivial Polyakov loop, ``dissolve'' in $n$ monopole constituents 
for large enough instanton scale parameters. We discuss recent 
results for these caloron solutions and their fermion zero-modes, 
as well as the implications for lattice studies and comment on 
the possible influence of the constituent monopoles on the 
instanton size distribution.
\end{abstract}
\maketitle
\section{Introduction}
Calorons for SU($n$), instantons at finite temperature, have $n$ 
monopoles as constituents~\cite{Nahm,Lee,KrvB} each with a mass 
$8\pi^2\nu_m/\beta$, where $\nu_m\!\equiv\!\mu_{m+1}\!-\!\mu_m$, 
defined in terms of the Polyakov loop at spatial infinity (the 
so-called holonomy) through a global gauge rotation $g$ and its 
(ordered) eigenvalues, 
\beqa
&&\pl=g\exp[2\pi i\,{\rm diag}(\mu_1,\ldots,\mu_n)]g^\dagger,\\
&&\mu_1\leq\ldots\leq\mu_n\leq\mu_{n+1}\!\equiv\!1\!+\!\mu_1,
~\sum_{m=1}^n\mu_m\!=\!0.\nonumber
\eeqa
Using the classical scale invariance we may set $\beta=1$ throughout. 
The charge one SU($n$) action density has a simple form~\cite{KrvB},
\beqa
&&\hskip-6mm\Tr F_{\mu\nu}^{\,2}(x)\!=\!\partial_\mu^2\partial_\nu^2
\log\left[\half\tr(\cA_n\cdots \cA_1)-\cos(2\pi t)\right],\nonumber\\
&&\hskip-6mm\cA_m\equiv\frac{1}{r_m}\left(\!\!\!\bea{cc}r_m\!\!&|\vec 
y_m\!\!-\!\vec y_{m+1}|\\0\!\!&r_{m+1}\eea\!\!\!\right)\left(\!\!\!
\bea{cc}c_m\!\!&s_m\\s_m\!\!&c_m\eea\!\!\!\right),
\eeqa
with $r_m\!=\!|\vec x\!-\!\vec y_m|$ the center of mass radius of the 
$m^{\rm th}$ constituent monopole, $c_m\!\equiv\!\cosh(2\pi\nu_m r_m)$, 
$s_m\!\equiv\!\sinh(2\pi\nu_m r_m)$, $r_{n+1}\!\equiv\! r_1$ and 
$\vec y_{n+1}\!\equiv\!\vec y_1$. 

\begin{figure}[bht]
\vspace{1.5cm}
\includegraphics{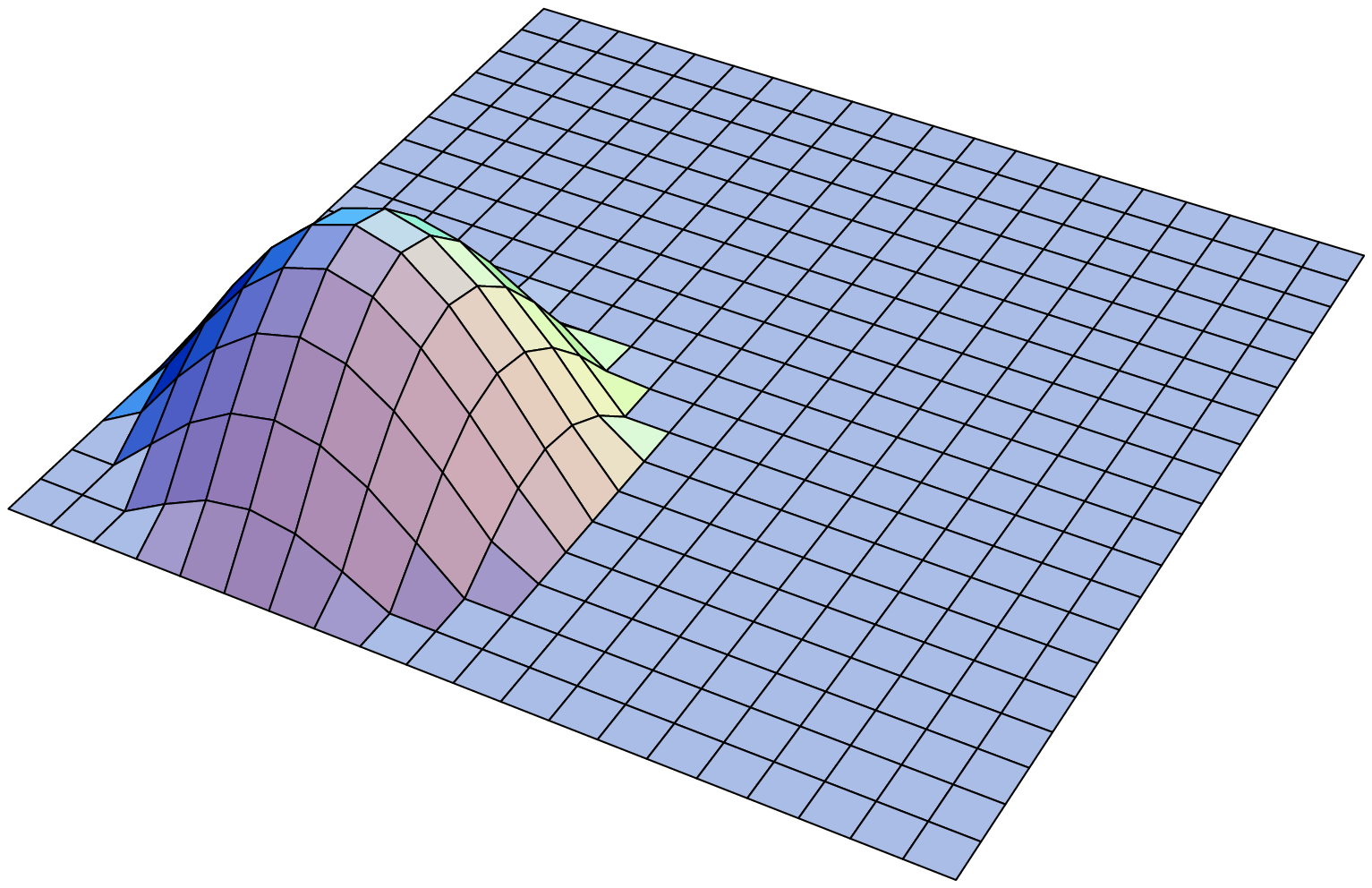}
\includegraphics{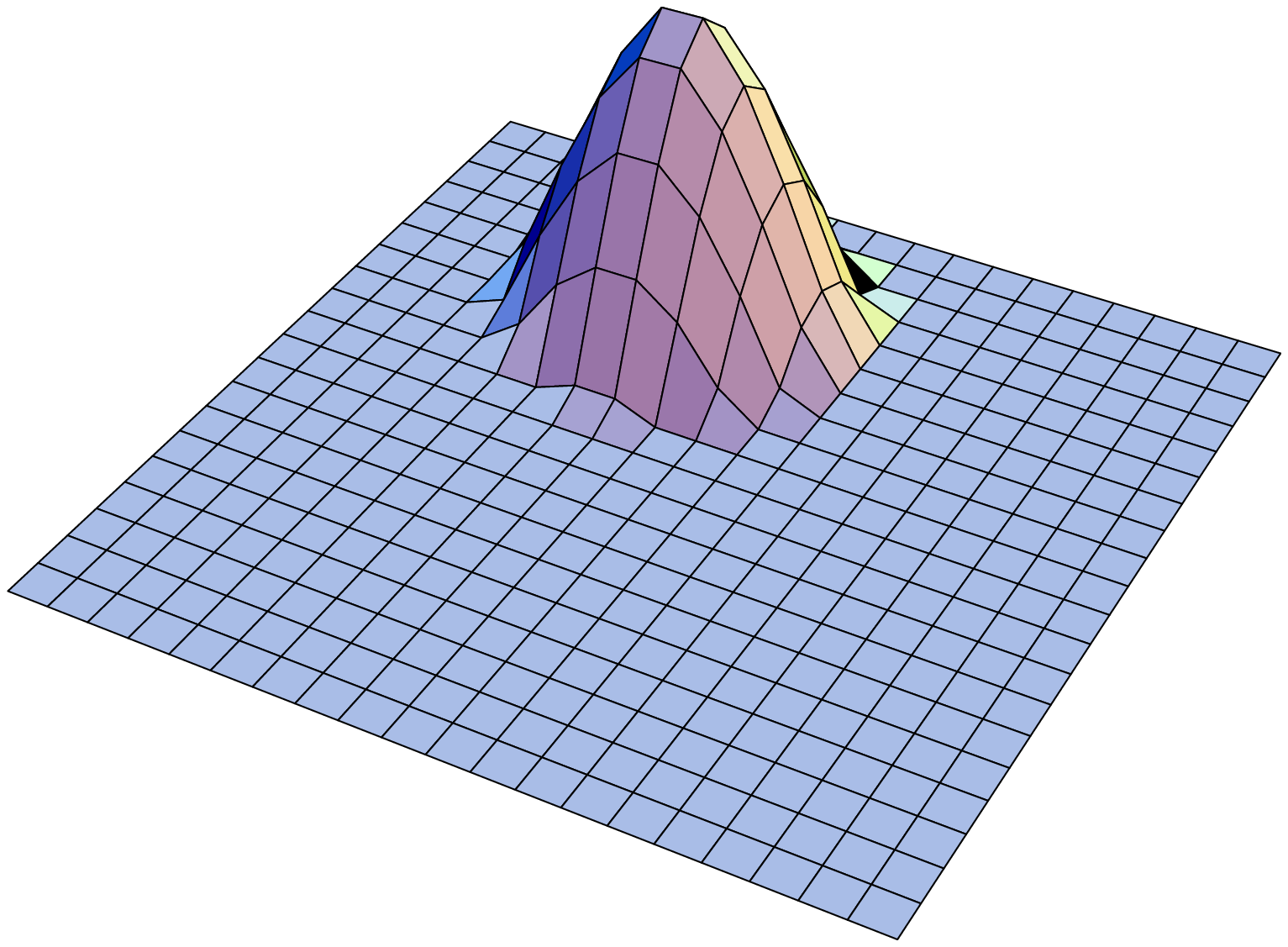}
\includegraphics{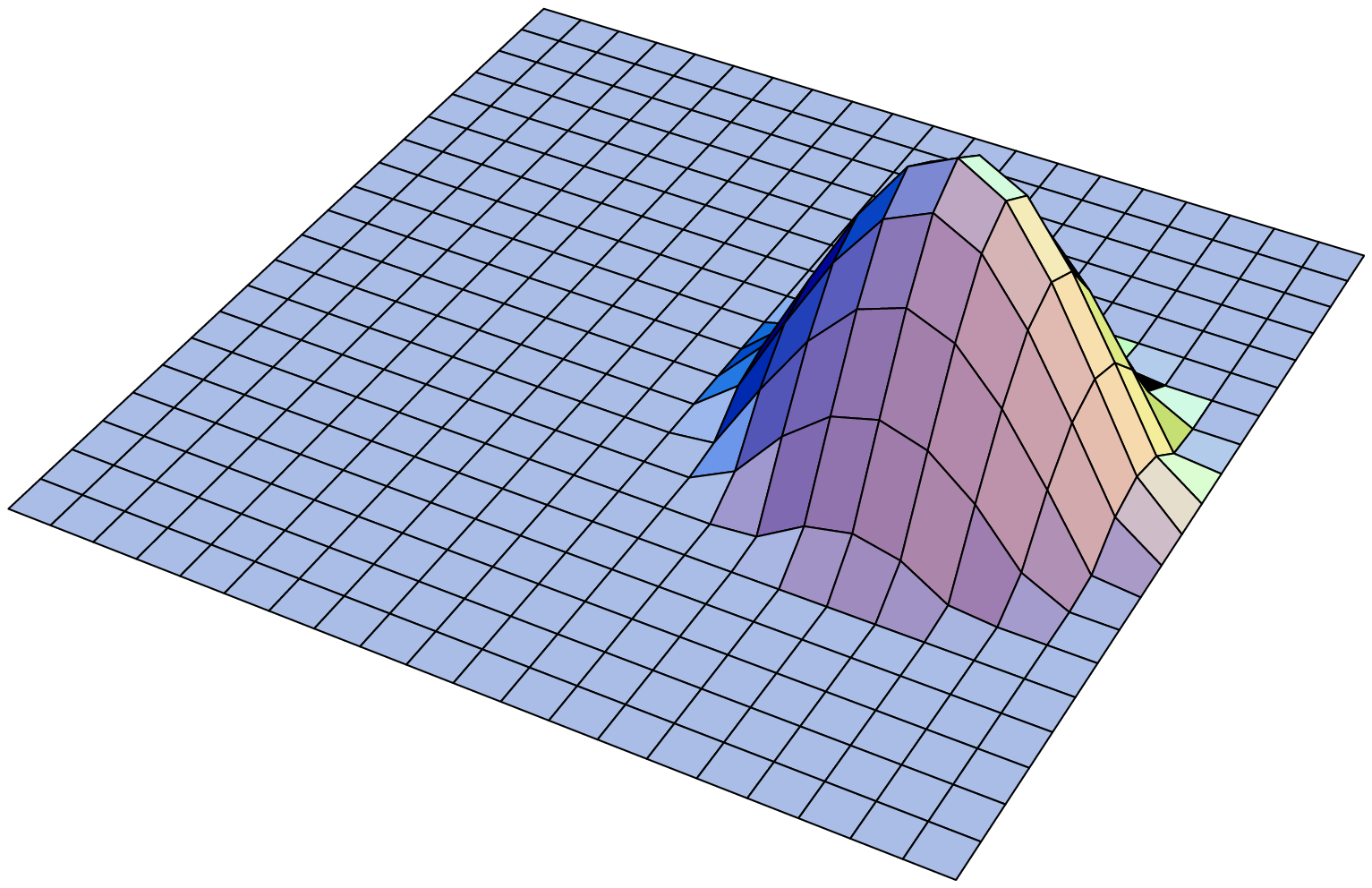}
\caption{Logarithm of zero-mode densities with $z=4/5,1/2,1/5$ (from left 
to right) for an SU(3) caloron with $(\mu_1,\mu_2,\mu_3)=(-17,-2,19)/60$.}
\label{fig:1}
\end{figure}

The fermion zero-mode is localized~\cite{ZM2,ZMN} on the $m^{\rm th}$ 
constituent monopole, if $\mu_m\!<\!z\!<\!\mu_{m+1}$ specifies its boundary 
condition $\Psi_z(t+\beta,\vec x)\!=\exp(-2\pi iz)\pl\Psi_z(t,\vec x)$ in the 
so-called algebraic gauge, where $A_\mu(t+\beta,\vec x)=\pl A_\mu(t,\vec x)
{\cal P}_\infty^{-1}$. ``Cycling'' through the values of $z$ gives a distinct 
signature, see e.g. Fig.~\ref{fig:1}, through which one can identify 
well-dissociated calorons, as observed for SU(2)~\cite{IMMV} with, and 
for SU(3)~\cite{Gatt} without cooling.

\section{Progress at higher charge}
The analytic formalism can be generalized to higher charge calorons 
($Q\!>\!1$), although explicit results are harder to get at. In this 
section we report on work by three of us (FB, DN and PvB)~\cite{BNvB}. 

Periodically repeating $Q$ instantons (up to a colour rotation with $\pl$) 
within each time-layer $t\in[p, p+1]$, using $p$ as a Fourier index, maps the 
appropriate ADHM data~\cite{ADHM} to a self-dual $U(Q)$ gauge field $\hat 
A_\mu(z)$. This is defined on a circle parameterized by $z\in[0,1]$, with $n$ 
well-defined singularities at $z=\mu_j$, in accordance with the Nahm 
transformation~\cite{Nahm}. When $\hat A(z)$ is constant for $z\in[\mu_j,
\mu_{j+1}]$ (always true for $Q=1$), its eigenvalues are directly related to 
the constituent locations, all of the same type with a mass $8\pi^2\nu_j$.

\begin{figure}[bht]
\vspace{2.3cm}
\includegraphics{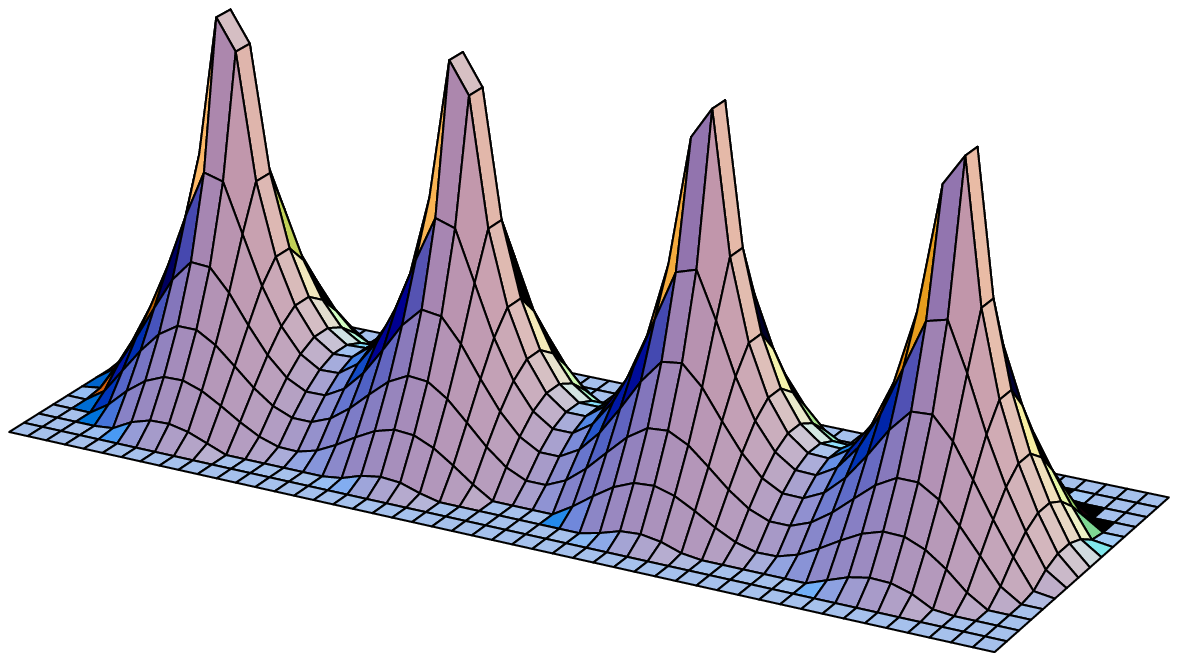}
\includegraphics{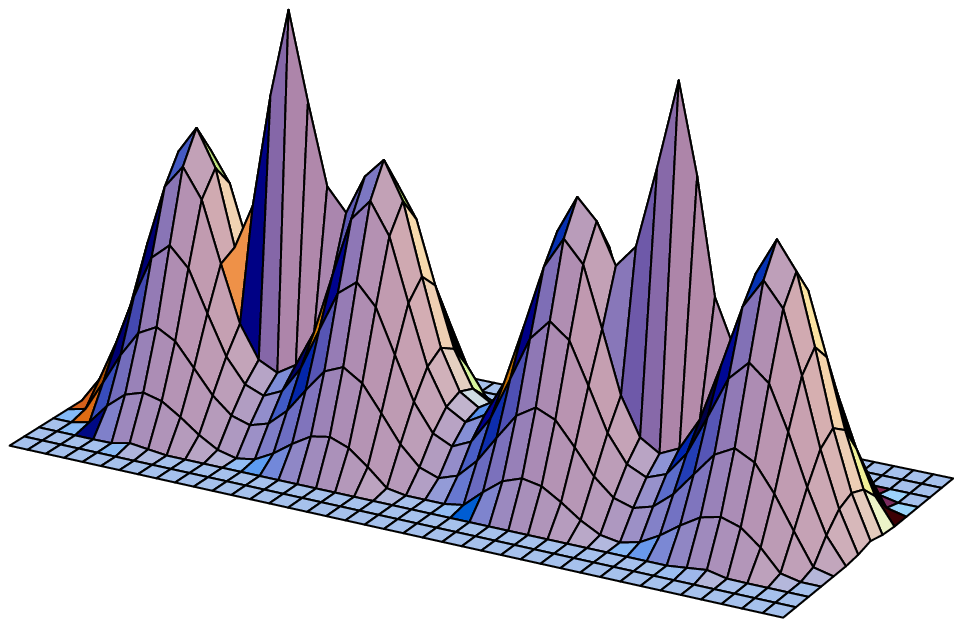}
\includegraphics{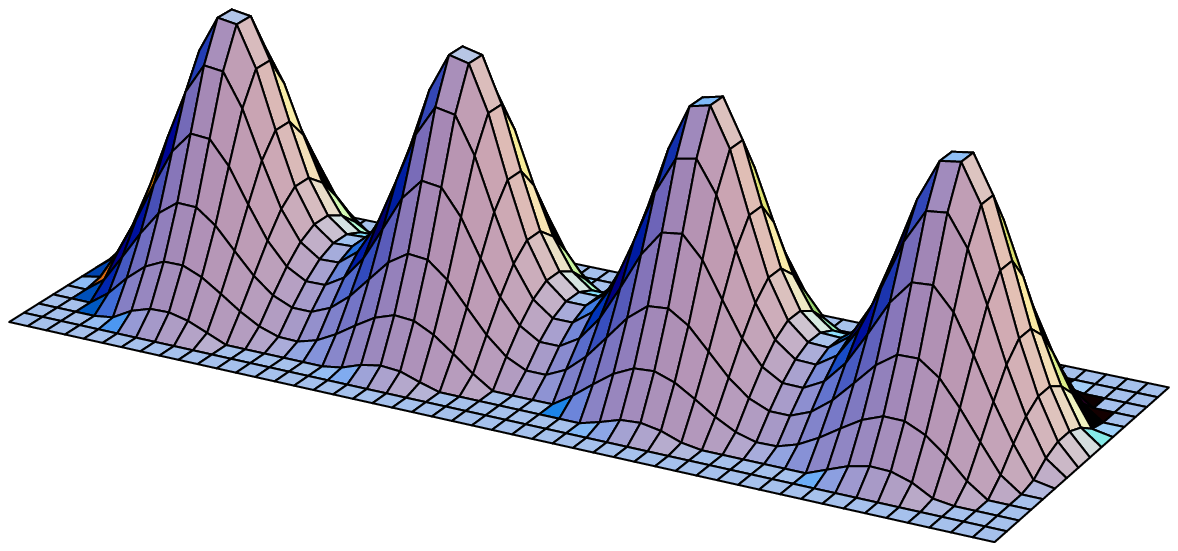}
\includegraphics{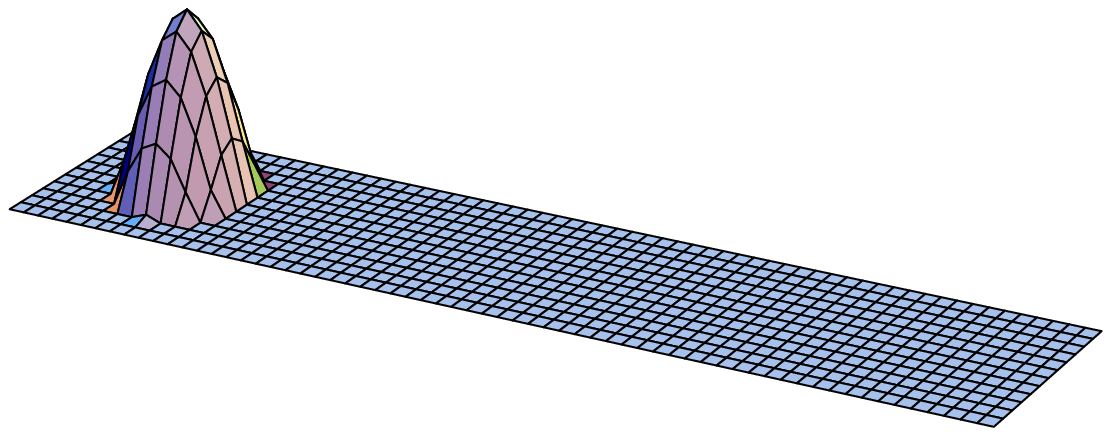}
\caption{Logarithm of action densities for the sum ansatz (top-left), 
high temperature limit and exact (right) solutions of an SU(2) charge 2 
caloron. Bottom-left corner shows one of the two anti-periodic 
($z\!=\!\half$) zero-mode densities.}\label{fig:2}
\end{figure}

The motivation for these studies has been to understand to which degree the 
monopole constituents can be seen (outside their non-abelian cores) as 
independent point-like Dirac monopoles. Approximate superpositions of 
calorons, e.g. using the sum ansatz~\cite{ScSh}, suffer from subtle effects 
that lead to finite energy remnants of the Dirac string as illustrated in 
Fig.~\ref{fig:2}. Fortunately this is an artefact of the approximation, as 
can be seen from a class of exact axially symmetric solutions, for which 
$\hat A(z)$ is piecewise constant. It does, however, make model building 
on the basis of these constituent monopoles intricate. 

In the high temperature limit the non-abelian cores of the monopoles shrink 
to zero size. As long as $z\neq\mu_j$ the mass of the fermions becomes 
infinite in this limit and the zero-modes are entirely localized to the 
non-abelian cores of the constituent monopoles, cmp. Fig.~\ref{fig:2}. The 
fermion density, summed over all zero-modes, is given by $\sum_a|\Psi^{(a)}
(x;z)|^2 =-(2\pi)^{-2}\partial_\mu^2\Tr\hat f_x(z,z)$, where $\hat f_x(z,z')$ 
is the Green's function of the ADHM-Nahm construction and can be computed 
using impurity scattering methods. Many simplifications occur in the 
high temperature limit, where the density for $\mu_m\!<\!z\!<\!\mu_{m+1}$ 
can be written as $-\partial_i^2\cV_m(\vec x)$. This function can therefore 
be used to trace the possible extended nature of constituent monopole sources 
(independence of $z$ implies it is a highly non-trivial conserved quantity 
of the Nahm equations). For SU(2) and $Q\!=\!2$ we found
\beq
\cV_m(\vec x)=\frac{1}{2\pi|\vec x|}+\frac{\cD}{4\pi^2}\int_{r<\cD}\hskip-5mm
drd\varphi\frac{\partial_r|\vec x-r\vec y(\varphi)|^{-1}}{\sqrt{\cD^2-r^2}},
\eeq
where $\vec y(\varphi)\!=\!(\sqrt{1-\k^2}\cos\varphi,0,\sin\varphi)$, up to an 
arbitrary coordinate shift and rotation, in terms of a scale $\cD$ and shape 
parameter $\k$ to characterize {\em arbitrary} SU(2) charge 2 solutions of 
$\hat A(z)$.

This represents an extended structure with the core restricted to a disk 
bounded by an ellipse with minor axes $2\cD\sqrt{1\!-\!\k^2}$ and major axes 
$2\cD$. But for large $\cD$ this can be shown to approach two point-like 
monopoles ($\k\!\to\!1$) separated by $2\cD$. Axially symmetric solutions 
are always point-like ($\k\!=\!1$), with unlike-charges alternating.

\section{Truly observing constituents?}
We return to interpreting the recent lattice data of Ref.~\cite{Gatt},
based on using the fermion zero-modes as a filter, restricting to Q=1 by 
insisting there is only one ``exact'' zero-mode. In the confined phase a 
reasonable fraction of configurations was found with one lump in the 
zero-mode density, but at different locations as a function of $z$.
This fitted well the expected behaviour of a charge one SU(3) caloron, see 
Fig.~\ref{fig:1}. There will typically be near zero-modes as well, 
revealing the presence of Q anti- and Q+1 instantons. To rule out that 
the zero-mode could be hopping between the Q+1 instantons, rather than 
between the constituent monopoles of a single caloron, two of us (MGP 
and PvB) computed the mixing of zero-modes as a function of $z$. 

For simplicity we consider SU(2), adding a charge 1 anti-instanton and a 
charge 2 instanton in the algebraic gauge with (the same) arbitrary holonomy, 
i.e. using the sum ansatz~\cite{ScSh}. In isolation the charge 2 instanton has 
two exact zero-modes, and only one of them can survive: the one associated 
to the zero (left) eigenvector of the 2x1 overlap matrix~\cite{ScSh} 
(using Weyl fermions)
\beqa
O_{IA}^{ij}(z)=&\int d^4x\Psi^{(i)}_{(I)}(x;z)^\dagger\sigma_\mu D_\mu
\Psi^{(j)}_{(A)}(x;z)\nonumber\\=&\!\!\!-\int d^4x\Psi^{(i)}_{(I)}
(x;z)^\dagger\sigma_\mu\partial_\mu\Psi^{(j)}_{(A)}(x;z),
\eeqa
where $\sigma_\mu=(\Eins_2,i\vec\tau)$ and the equation of motion for the 
zero-modes was used. We now assume that the charge 2 instanton is made up 
of two sufficiently separated instantons, in which case $\Psi^{(i)}_{(I)}$ 
to a good approximation is equal to the charge 1 zero-mode associated with 
the $i^{\rm th}$ instanton. 

The exact form of the SU(2) Q=1 zero-mode with $\pl=\exp(2\pi i\omega\tau_3)$
has been given before~\cite{ZM2}. The anti-instanton and its zero-mode are 
obtained by time reversal, which changes the periodicity to 
${\cal P}_\infty^\dagger\exp(2\pi iz)$, to be corrected by a gauge rotation 
$i\tau_2$ and changing the sign of $z$. This ensures the overlap integral is 
periodic. In terms of the scale parameter $\rho$ (determining the distance
$\pi\rho^2/\beta$ between the constituents), arbitrary location $a_\mu$ and 
orientation $R_{ij}=\half\tr(U\tau_i U^\dagger\tau_j)$ (with $U\!\in$ SU(2) 
the associated spin rotation), we find \vskip-1mm\noindent
\beqa
&&\hskip-6mm O_{IA}(z)\!=\!-\frac{\rho_{(I)}\rho_{(A)}}{4\pi^2}\!
\int\!d^4x\left(K(z)+K^*(-z)\right),\\
&&\hskip-6mm K(z)\!\equiv\!\psi_{(I)}^\dagger(x^{(I)};z)U^\dagger_{(I)}
\sigma_\mu\partial_\mu U_{(A)}\psi_{(A)}(x^{(A)};z),\nonumber
\eeqa
where, $x^{(I,A)}=(t-a^{(I,A)}_0,R^{(I,A)}(\vec x-\vec a^{(I,A)}))$ and
\beqa
&&\hskip-5mm\psi_{(I)}(x;z)\equiv\phi_{(I)}^\half(x)\pmatrix{\partial_2+
i\partial_1\cr\partial_0-i\partial_3\cr}\hat f^{(I)}_x(\omega,z),\nonumber\\
&&\hskip-5mm\psi_{(A)}(x;z)\equiv\phi_{(A)}^\half(x)\pmatrix{\partial_0-i
\partial_3\cr\partial_2-i\partial_1\cr}\hat f^{(A)}_x(\omega,z).
\eeqa
\vskip-1mm\noindent
Explicit expressions for $\phi(x)$ and $\hat f_x(\omega,z)$ can be found 
elsewhere~\cite{KrvB,ZM2,ZMN}. 

For generic values of the parameters the overlap is complex and to go 
through zero (to maximize localization) requires fine-tuning of two 
parameters. We will assume here that, for each of the two instantons and 
the anti-instanton, both constituent monopoles are approximately at the same 
location ($\pi\rho^2/\beta$ small). Typically then, if one instanton is 
sufficiently further away from the anti-instanton their overlap will be 
exponentially small and the zero mode will localize on that instanton. 
This happens generically for any $z$ except for two small intervals 
(proportional to $\beta^{-1}$) around $z=\pm\omega$ where the (anti-)instanton 
zero mode becomes delocalized and the overlap is not guaranteed to be 
small. One could, in principle, observe in these two intervals a 
significant mixing with the zero mode of the second instanton. Notice 
however that, except for trivial holonomy where the two intervals merge 
into one (at $z=0$ or $\half$), this effect leads to double the expected 
number of hoppings. Therefore it could only possibly mimic the behaviour of a 
{\it single} caloron with trivial holonomy. We can then conclude that 
the full signature of a single SU(2) caloron with {\em non-trivial} 
holonomy is unlikely to be emulated by the effect of hopping between 
two instantons, and we expect the same to hold for SU(3).

\section{Discussions}
Cooling studies at a low temperature, rather than just below the deconfinement 
temperature, have been reported as well~\cite{IMMV}. Constituents at 
non-trivial holonomy could still be identified by points where two
eigenvalues of the Polyakov loop coincide, but they revealed no individual 
lumps. This may explain why before no constituents were found. It is also 
interesting to point out that for $\rho\!>\!\beta$ the scale parameter plays 
an entirely different role and the measure for integration over the moduli 
space should be formulated in terms of constituent monopoles of fixed size 
(at fixed temperature). This should remove the infrared divergence of the scale 
integration, although it is unlikely a semiclassical picture can be used here.

\section*{Acknowledgements}
We thank Michael M\"uller-Preussker, Michael Ilgenfritz and Christof 
Gattringer for discussions. FB is supported by FOM and MGP by MCYT.

\end{document}